\documentclass[aps,twocolumn]{revtex4-1}

\newcommand{\ben}{\begin{eqnarray}}
\newcommand{\een}{\end{eqnarray}}
\begin{document}

\title{Primordial scalar perturbations in tachyonic power-law inflation}%
\author{Rudinei C. de Souza}%
\author{Gilberto M. Kremer}%
\affiliation{Departamento de F\'isica, Universidade Federal do Paran\'a, Brazil}

\begin{abstract}
In this work we determine the power spectrum of the gravitational potential of the primordial fluctuations for an inflationary model whose \emph{inflaton} is a non-canonical scalar field of the tachyon-type. The respective background field equations for an inverse-square potential produce a power-law inflation, and it is explicitly shown that for such a potential the power spectrum tends to be scale-independent for highly accelerated regimes in the inflationary expansion.
\end{abstract}

\maketitle

\section{About the model} The tachyon field has received some attention in cosmology, being applied to produce an accelerated expansion of the universe. It can play an important role in inflationary models \cite{A, B, C, D, E, E1, F, G} as well as in the present accelerated expansion, simulating the effect of the dark energy \cite{C, 1, 2, 3, 4, 5}. The inverse-square potential we consider in our model was firstly proposed in \cite{A} and it can promote a power-law accelerated expansion. The analysis of the dynamics of the respective potential was performed in references \cite{C, 1, 2, 3, 4}. The work \cite{C} investigated the inverse power-law potentials in a general way and asymptotic computations of the density perturbations were performed. But the details concerning to the tachyonic power-law inflation are not studied. In the reference \cite{Steer} it is done a comparison between the standard inflation and the tachyonic one. The power spectrum of the perturbations is calculated through a slow roll formalism and applied to some typical potentials, including  an inverse power-law potential, but not the inverse-square one. The subject of the present paper is to analyse the primordial perturbations through an explicit calculation of the power spectrum of the gravitational potential. Solutions in terms of Bessel functions came out, generating power spectrums for the long wave-length limit which depend on the exponent of the wave-number. In one of them, valid for integer indices of the Bessel functions, the power spectrum is scale-dependent. The other one, valid for non-integer indices, can produce a \emph{quasi} scale-independent power spectrum if the indices tend to the value 1/2. This condition implies that the flatness of the power spectrum holds if the cosmic expansion is highly accelerated, which is in line with the background solution, i.e., a tachyonic power-law inflation generated by an inverse-square potential.

\section{Background field equations}
The model we consider here is described by the following action
\begin{equation}
S=\int d^4x\sqrt{-g}\left\{\frac{R}{2}+{V}{\sqrt{1-g^{\mu\nu}\partial_\nu\varphi\partial_\nu\varphi}}\right\},\label{action}
\end{equation}
where $R$ is the Ricci scalar and $\varphi$ is a non-canonical scalar field of the tachyon-type with the self-interaction potential $V(\varphi)$. For a flat Friedmann-Robertson-Walker (F-R-W) metric, the background energy density and pressure of the tachyon condensate are
\begin{equation}
\rho_0=\frac{V}{\sqrt{1-\dot\varphi_0^2}}, \qquad p_0=-V\sqrt{1-\dot\varphi_0^2},\label{de_pressure}
\end{equation}
respectively. The corresponding Friedmann equation and time evolution equation of the field read
\begin{equation}
H^2=\frac{V}{3\sqrt{1-\dot\varphi_0^2}}, \qquad \frac{\ddot\varphi_0}{1-\dot\varphi_0^2}
+3H\dot\varphi_0+\frac{1}{V}\frac{dV}{d\varphi_0}=0,\label{fequations}
\end{equation}
respectively, where $H=\dot a/a$ is the Hubble parameter, $a$ is the scale factor and the dot denotes derivative with respect to time.

By considering an inverse-square potential
\begin{equation}
V=\frac{\lambda}{\varphi_0^{2}}, \label{potential}
\end{equation}
with $\lambda$ being a constant, the above system is satisfied by the solution
\begin{equation}
a(t)=a_0(t+c_1)^n, \qquad \varphi_0(t)=\sqrt{\frac{2}{3n}}(t+c_1).\label{fe_solution}
\end{equation}
Here $a_0, c_1$ and $n$ are constants and the following relation between $n$ and $\lambda$ holds
\begin{equation}
n=\frac{1}{6}\left(2+\sqrt{4+9\lambda^2}\right).\label{n_lambda}
\end{equation}
Such a model describes a power-law inflation if $\lambda>2/\sqrt{3}$. The speed of sound corresponding to this solution reads
\begin{equation}
 c_s=\sqrt{\frac{3n-2}{3n}}.
\end{equation}

\section{Cosmological perturbations} Since the fluid of the model does not present anisotropic stress, the small scalar perturbations in the flat F-R-W line element in the longitudinal gauge \cite{Bardeen, Mukhanov, Mukhanov1} are represented by
\begin{equation}
ds^2=a^2(\eta)\left[(1+2\Phi)d\eta^2-(1-2\Phi)\delta_{ij}dx^idx^j\right]. \label{pline}
\end{equation}
Above $\eta$ is the conformal time, defined as $d\eta=dt/a$, and $\Phi$ is the gravitational potential of the perturbation.

Introducing the Mukhanov-Sasaki variables \cite{Mukhanov, Mukhanov1}
 \begin{equation}
u=\frac{2\Phi}{\sqrt{\rho_0+p_0}}, \qquad \theta=\frac{1}{\sqrt{3}}\left(\frac{1}{a}\right)\left(1+\frac{p_0}{\rho_0}\right)^{-\frac{1}{2}},\label{ms_variables}
\end{equation}
the differential equation that describes the perturbations reads
 \begin{equation}
u''-c_s^2\nabla^2u-\frac{\theta''}{\theta}u=0,\label{pequation}
\end{equation}
where the prime denotes derivative with respect to the conformal time.

Rewriting (\ref{pequation}) such that $u$ is a function of the scale factor, we can set it in the form
 \begin{equation}
a^2\frac{d^2u}{da^2}+a\left(2+\frac{\dot H}{H^2}\right)\frac{du}{da}-\frac{c_s^2}{a^2H^2}\nabla^2u+\frac{\dot H}{H^2}u=0.\label{pequation1}
\end{equation}
If we apply the background solution to the above equation and consider a plane-wave solution in the form $u_\textbf{k}e^{-i\textbf{k.x}}$, one has
 \begin{equation}
a^2\frac{d^2u_\textbf{k}}{da^2}+\left(2p+1\right)a\frac{du_\textbf{k}}{da}+
\left[k^2\alpha^2a^{-4p}-\frac{1}{n}\right]u_\textbf{k}=0,\label{pequation2}
\end{equation}
where
 \begin{equation}
p=\frac{n-1}{2n}, \qquad \alpha^2=\frac{c_s^2}{a_0^{2/n}n^2}.\label{constants}
\end{equation}

Equation (\ref{pequation2}) is one of the forms of the Bessel equation, whose solution can be written as
 \begin{equation}
u_\textbf{k}=a^{-p}\left[A_\textbf{k}J_q\left(\beta k a^{-2p}\right)+B_\textbf{k}Y_q\left(\beta k a^{-2p}\right)\right],\label{psolution}
\end{equation}
which is valid for integer $q$. For non integer $q$, its form is
 \begin{equation}
u_\textbf{k}=a^{-p}\left[A_\textbf{k}J_q\left(\beta k a^{-2p}\right)+B_\textbf{k}J_{-q}\left(\beta k a^{-2p}\right)\right].\label{psolution1}
\end{equation}
Above $A_\textbf{k}$ and $B_\textbf{k}$ are constants and
\begin{equation}
q=\frac{n+1}{2(n-1)}, \qquad \beta=\frac{n\alpha}{(n-1)}.\label{constants1}
\end{equation}

Firstly, let us calculate the asymptotic forms of the solution for integer $q$. For short-wavelength perturbations, which satisfy $ka^{-2p}\gg1$, one obtains by using the asymptotic forms of the Bessel functions for a large argument in (\ref{psolution})
\begin{equation}
u_\textbf{k}\simeq C_{\textbf{k}}e^{i\beta k (a^{-2p}-a^{-2p}_i)}.\label{MS_asymp1}
\end{equation}
We have set $B_\textbf{k}=iA_\textbf{k}$ and defined
\begin{equation}
C_{\textbf{k}}=\sqrt{\frac{2}{\pi\beta k}}A_\textbf{k}e^{-i\xi} e^{i\beta k a^{-2p}_i}, \qquad \xi=\frac{\pi n}{2(n-1)},\label{constants2}
\end{equation}
being $a_i=a(t_i)$ or $a_i=a(\eta_i)$, where $t_i$ and $\eta_i$ stand for an initial instant. These redefinitions for the constants will facilitate the fixing of the initial conditions later. For long-wavelength perturbations, when $ka^{-2p}\ll1$, one has by employing the asymptotic forms of the Bessel functions for a small argument in (\ref{psolution})
 \begin{equation}
u_\textbf{k}\simeq C_{1\textbf{k}}a^{\frac{1}{n}}+C_{2\textbf{k}}a^{-1}.\label{MS_asymp2}
\end{equation}
 The constants were redefined as
 \begin{equation}
 C_{1\textbf{k}}=-\frac{i}{\pi}A_\textbf{k}\Gamma(q)\left(\frac{\beta k}{2}\right)^{-q}, \quad C_{2\textbf{k}}=\frac{A_\textbf{k}}{\Gamma(1+q)}\left(\frac{\beta k}{2}\right)^{q}.\label{constants3}
\end{equation}

Now, proceeding in the same way as we did above, from equation (\ref{psolution1}) one calculates the asymptotic forms of the solution for non integer $q$. For the limit of short-wavelength perturbations, we get by setting $B_\textbf{k}=iA_\textbf{k}$ for convenience
\begin{equation}
u_\textbf{k}\simeq D_{\textbf{k}}\left[\cos{\left(\beta ka^{-2p}-\xi\right)}+i\cos{\left(\beta ka^{-2p}-\xi+\pi q\right)}\right],\label{MS_asymp3}
\end{equation}
where
\begin{equation}
D_{\textbf{k}}=\sqrt{\frac{2}{\pi\beta k}}A_{\textbf{k}}.\label{constants4}
\end{equation}
For the limit of long-wavelength perturbations, we have
\begin{equation}
u_\textbf{k}\simeq D_{1\textbf{k}}a^{\frac{1}{n}}+D_{2\textbf{k}}a^{-1},\label{MS_asymp4}
\end{equation}
with
\begin{equation}
D_{1\textbf{k}}=-\frac{iA_{\textbf{k}}}{\Gamma(1-q)}\left(\frac{\beta k}{2}\right)^{-q}, \quad D_{2\textbf{k}}=\frac{A_{\textbf{k}}}
{\Gamma(1+q)}\left(\frac{\beta k}{2}\right)^{q}.\label{constants5}
\end{equation}

\section{Power spectrum for integer $q$}  According to the current inflationary theory, the primordial quantum fluctuations generated the seeds for the large scale structures. Thus the primordial power spectrum must be consistent with the minimal fluctuations of energy allowed by quantum mechanics, i.e., we have to preserve the fluctuations of the vacuum. Such a requirement is satisfied if the initial conditions for the variable $u_{\textbf{k}}$ present the forms \cite{Mukhanov, Mukhanov1}
\begin{equation}
u_{\textbf{k}i}=-\frac{i}{\sqrt{c_s}}k^{-\frac{3}{2}}, \qquad u'_{\textbf{k}i}={\sqrt{c_s}}k^{-\frac{1}{2}}.\label{IC}
\end{equation}

To satisfy these initial conditions, from (\ref{MS_asymp1}) we determine that $C_\textbf{k}$ must have the following value
\begin{equation}
C_{\textbf{k}}=-i\frac{a_0^{{1}/{n}}\sqrt{\lambda}}{2c_sk^{\frac{3}{2}}}.\label{fixedC}
\end{equation}
So, the $u_\textbf{k}$ for short wave-length perturbations that satisfy the minimum fluctuations of energy read
\begin{equation}
u_{\textbf{k}}\simeq-\frac{i}{\sqrt{c_s}k^{\frac{3}{2}}}e^{i\beta k (a^{-2p}-a^{-2p}_i)}.
\end{equation}
The corresponding power spectrum is
\begin{equation}
\delta_{\Phi}^2(k, a)\simeq
\frac{\lambda}{16\pi^2c_s^2}\left(\frac{a}{a_0}\right)^{-\frac{2}{n}},
\end{equation}
which is scale-independent.

From result (\ref{fixedC}), through the use of $(\ref{constants2})_1$ and (\ref{constants3}), we determine $A_\textbf{k}$ and the constants of solution (\ref{MS_asymp2}). Since $k^2\ll a^{-2p}$ for this solution, one can keeps only its first term, obtaining the respective $u_\textbf{k}$
\begin{equation}
u_{\textbf{k}}\simeq-\frac{\Gamma(q)}{\sqrt{\pi c_s}k^{1+q}}\left(\frac{\beta}{2}\right)^{\frac{1}{2}-q}{a}^{\frac{1}{n}}.
\end{equation}
This furnishes the following power spectrum
\begin{equation}
\delta_{\Phi}^2(k, a)\simeq\frac{\lambda a_0^{{2}/{n}}\Gamma(q)^2}{16\pi^3c_s^2}\left(\frac{\beta}{2}\right)^{1-2q}k^{1-2q},
\end{equation}
valid for long wave-length perturbations. Such a power spectrum is scale-dependent for all the $q$. Note that this result is valid only for integer $q$.

\section{Power spectrum for non integer $q$}  For this case, by using (\ref{MS_asymp3}), the initial condition (\ref{IC}) is satisfied if $D_\textbf{k}$ has the form
\begin{equation}
D_{\textbf{k}}=-\frac{ia_0^{{1}/{n}}\sqrt{\lambda}}{2c_sk^{\frac{3}{2}}}
[\cos{\left(\beta ka_i^r-\xi\right)}+i\cos{\left(\beta ka_i^r-\xi+\pi q\right)}]^{-1}.\label{fixedD}
\end{equation}
Thus the adequate $u_\textbf{k}$ for the limit of short wave-length perturbations is
\begin{equation}
u_{\textbf{k}}\simeq-\frac{i}{\sqrt{c_s}k^{\frac{3}{2}}}\left[\frac{\cos{\left(\beta ka^r-\xi\right)}+i\cos{\left(\beta ka^r-\xi+\pi q\right)}}{\cos{\left(\beta ka_i^r-\xi\right)}+i\cos{\left(\beta ka_i^r-\xi+\pi q\right)}}\right].
\end{equation}

The power spectrum for this limit is
\ben
&\delta_{\Phi}^2(k, a)\simeq\frac{\lambda}{16\pi^2c_s^2}\nonumber\\
&\times\left[\frac{\cos^2{\left(\beta ka^r-\xi+\pi q\right)}+\cos^2{\left(\beta ka^r-\xi\right)}}{\cos^2{\left(\beta ka_i^r-\xi+\pi q\right)}+\cos^2{\left(\beta ka_i^r-\xi\right)}}\right]\left(\frac{a}{a_0}\right)^{-\frac{2}{n}}.\label{PE1}
\een

From result (\ref{fixedD}), and (\ref{constants4}) through (\ref{constants5}), the asymptotic form (\ref{MS_asymp4}), by despising its second term, renders
\ben
&u_{\textbf{k}}\simeq-\frac{i\sqrt{2\pi}\left({\beta}/{2}\right)^{\frac{1}{2}-q}k^{-1-q}}{2\sqrt{c_s}\ \Gamma(1-q)}\nonumber\\
&\times\left[\cos{\left(\beta ka_i^r-\xi\right)}+i\cos{\left(\beta ka_i^r-\xi+\pi q\right)}\right]^{-1}{a}^{\frac{1}{n}}.
\een
Hence the respective power spectrum, valid for long wave-length perturbations, reads
\ben
&\delta_{\Phi}^2(k, a)\simeq\frac{\lambda a_0^{{2}/{n}}\left(\beta/2\right)^{1-2q}}{32\pi c_s^2\Gamma^2(1-q)}\nonumber\\
&\times\left[\cos^2{\left(\beta ka_i^r-\xi+\pi q\right)}+\cos^2{\left(\beta ka_i^r-\xi\right)}\right]^{-1}{k^{1-2q}}.\label{PE2}
\een

Analysing the quantities (\ref{PE1}) and (\ref{PE2}), we can infer that they tend to be scale-independent when $q\rightarrow1/2$. On the other hand, from expressions $(\ref{fe_solution})_1$ and $(\ref{constants1})_1$ one has the following relations connecting $q$, $n$ and the scale factor
\begin{equation}
q=\frac{n+1}{2(n-1)}, \qquad a(t)=a_0(t+c_1)^n,
\end{equation}
which show that the situation $q\rightarrow1/2$ occurs when $n$ is sufficiently large. Note that this holds when the rate of the inflationary expansion is highly accelerated. Thus the present inflationary model produces a power spectrum which is progressively more scale-independent whenever the primordial expansion of the universe is more accelerated.

\section{Conclusion.}  In this work we have analyzed a tachyon field as the source of a power-law inflation generated by an inverse-square potential. Solutions in terms of Bessel functions were found for the gravitational potential of the perturbations. For integer indices of the Bessel functions it is not possible a flat power spectrum, but for non-integer indices there is a value, namely,  $q\rightarrow1/2$, for which the power spectrum turns out to be asymptotically flat, corresponding to a highly accelerated regime.

\end{document}